# 音乐中的时间箭头

—— 以可区分性以及唯一指向性为锚点重新审视音乐的时间结构

徐 起

**摘 要：** 文章由"时间箭头"这一广义的话题驱动，通过参考哲学（认识论）和科学（热力学）作为时间箭头词源意义上的上游，论述了时间箭头的两个具体性质：可区分性以及唯一指向性。文章以这两个性质为锚点，分别展开了对应的音乐命题并应用于具体的案例分析：对于可区分性，文章关注了音乐中的"复现"，从"诞生/重生"这一概念对切入，对巴赫的《圣诞节清唱剧》的定位试作新的解读；对于唯一指向性，文章论述了音乐中高点的后移，提出"AB-AAB左偏向模型"，将音乐的时间结构（例如曲式）看作动生生长过程的产物，并围此埋下一个有机主义（organicism）的伏笔。

**关键词：** 时间箭头；有机主义；申克分析；认识论；热力学；曲式；奏鸣曲式

## 引言：从聆听说开去

在音乐讨论中，如何"聆听"是一个永恒的话题：音乐始于聆听，也归于聆听。这一话题的有趣之处在于，虽然看上去不言自明的常识，却被持续探究，仿佛对这一话题的探究永无止境。例如，音乐学家杨燕迪谈及了"聆听"是如何在显而易见的同时，又大有文章：

> 就音乐评论而言，其中最重要的关键正在于——聆听、倾听并从中收获感动。听，这本是音乐艺术最本质的诉求，也是日常音乐活动中最常见的现象，所有音乐的接受者都必须聆听，而音乐也正是为听而存在，其中似乎无须再做理论。但是，究竟在音乐（作品和表演）中听什么？怎样听？听的美学真谛何在？具体到音乐评论者，又要关注听的什么方面，留意听到作品和表演中的什么要素？其中其实还有诸多疑难和学理问题有待澄清。[①]

虽然上文是从音乐评论的视角谈及聆听，但我们需要注意到：文中所提及的问题，不仅仅面向音乐评论者，而是值得所有音乐参与者（包括了听

众、评论者、演奏者、创作者、研究者）思考和探究。也就是说，聆听的重要性，对参与到音乐活动中的任何一方都成立。以笔者的（钢琴）演奏者身份为例，如何在演奏的时候仔细聆听自己是一个经久不衰的课题。一方面，根据演奏的定义，我们必然会在弹奏的时候听到自己的演奏。然而，"听到"不等于"聆听"；所以另一方面，学习如何主动高效地听自己，是所有演奏者需要穷尽一生探究的课题，以至于对"听"的讨论，在专业级别的演奏中被反复挖掘（我们甚至可以激进提出：水平越专业，注意力越会回归到聆听）。笔者的恩师任昭义在教学中强调的核心内容之一就是声音质量，以及为了追求声音质量而必须掌握的技能：带有想象力的聆听。见微而知著，但老师对聆听的重视并非孤例：在笔者求学的经历中，聆听是一个频繁浮现出的话题，在各种不同的上下文中从不同的角度被谈及。我们甚至可以试提出：个人乃至教学法的突破，主要来源之一就是对聆听的持续发掘。

为何"聆听"如同无穷尽的宝藏一般，值得被反复发掘和持续探究？笔者在此处无意尝试对这一值得投入一部专著，乃至一生来研究的问题进行解答，而是试着抛出一个主观且片面的命题，并

以此提供一个启发性的视角。我们可以试提出：在音乐中，之所以聆听如此重要，以至于被持续讨论，就是因为只有通过聆听，我们才能抓住音乐最动人心弦的特质，即通过声音动态的起伏变化，打动人心。那么从这个角度来说，聆听让我们更敏锐地感知到时间的流逝，以及它改变事物的能力。聆听音乐时的屏气凝神，让我们更加真切地与古人的感慨相共鸣：

惟草木之零落兮，恐美人之迟暮。②

于是，在曲终一刻，我们潸然泪下。如果我们非要追问这啜泣因何而起，它首先源自我们对音乐的聆听；进一步来说，透过聆听，我们的思绪被带向了时间的流逝性，如同潺潺溪流一般牵动着我们的心境。本文也在这里透过聆听这一话题，引出本文的焦点：时间箭头。

## 时间的显与隐

音乐是基于时间的艺术。更具体地说，音乐作品的呈现依赖演奏这一行为，而演奏的过程必然依托时间这一载体。也就是说，时间为音乐这一艺术形式的存在提供了最为基础的可能性。从诗意的角度，时间之于音乐，犹如阳光之于鲜花：唯有在时间中，音乐才得以如画卷一般展开(unfold)，正如含苞放的鲜花，向阳而生。

给出了时间对于音乐的重要性，一个很自然的跟进则是了解时间所具备的性质和特征。我们可以假设：如果音乐在时间中发生，那么时间的特性也会反映在音乐结构中。进一步说，时间的最典型特征，也会最直接影响音乐结构。时间这一容器(container)，如同无限延伸的画布，它的褶皱，决定着其中故事的走向发展。

时间的一个特殊性质在于它的指向性(directedness)：时间的流逝是单方向，不可逆(irreversible)的。在这里，我们借用物理学中约定俗成的术语"时间箭头"来指代时间的这一特殊性质。不仅仅是一个术语，"时间箭头"还是一个诗意的比喻：川流不息的历史就像河流，只朝着一个方向奔腾，像是在冥冥之中，在世界的剧本上，在万物的生死簿上，画着一个箭头在指引着时间

的流向。

音乐演奏者近水楼台，因为直接负责演奏这一行为，对音乐中时间的重要性尤为敏感。具体来说，则是对时间箭头这一时间所具有的独特性质敏感。以笔者所从事的钢琴专业为例，即使不给出"时间箭头"一词的具体定义，在音乐讨论中我们也实质上在反复提及这一概念，只是用到了不一样的词汇：音乐的流动性，方向，语气，倾向，起伏，这些在钢琴课上老生常谈的话题都是时间箭头的体现。比如，在教学实践中，不论是初学阶段的课，还是专业院校级别的大师班，我们都能经常听到"乐句往前走"抑或是"不要原地踏步"等近乎约定俗成的描述，不仅仅是因为我们容易一不小心就忽视音乐的流动性，更是因为它对有表现力的音乐演奏来说至关重要，所以值得反复强调。

我们甚至可以说，从演奏者的角度出发，音乐表演的感染力很大一部分就来自于对时间箭头的操控和扭曲。比如说，一个演奏者往往引以为傲的能力就是营造所谓的"魔力瞬间"(magical moment)。营造魔力瞬间的具体策略之一，就是巧妙利用我们称作 rubato 的弹性节奏来形成有表现力的乐句起伏。

进一步来说，在文艺作品中，如果我们可以扭曲时间的流逝(例如，让时光倒流，让人死而复生)，那么往往可以得到有表现力和戏剧张力的艺术效果。歌德笔下的浮士德，在和魔鬼做交易的时候，魔鬼怂恿他签下出卖灵魂的合同。而浮士德，则用一句匪夷所思的话来回应：

请向我展示在采摘前就腐烂的果实，以及每日发新芽的树木！③

这一回应之所以匪夷所思，除了答非所问的隐晦，还在于它的反直觉。我们的常识是，果实在采摘后腐烂，树木每年发青一次。而这段话，正是故意通过夸张离奇及反直觉的描述，来诗意地扭曲我们对时间的感知。这种时间上的扭曲渲染出了现实中的不可能，让我感受到了一个缤纷多彩，光怪陆离，超凡脱俗(transcendental)的世界。那个一切皆有可能的世界，正是浮士德所开出的条件。

我们不妨再考虑几个情景：

---

② 屈原：《离骚》。

③ Johann Wolfgang von Goethe, Faust, trans. Peter Salm, vol. 1, *Bantam Books*, 1985, p. 131.

·地面上打碎的杯子自动回到桌子上,并且拼好了自己

·死而复生

·时光倒流

这几个情景就像是电影倒放一样,会给我们带来困惑和不解,因为在现实中几乎不可能发生。但是,艺术作品往往就利用了困惑和不解的情绪,来达到打动人心的效果。这就像是一枚硬币的一体两面:困惑和不解在硬币的一面,而另一面就是震撼和感动。

颇为讽刺的是,尽管上文简述了时间箭头对音乐演奏乃至于艺术表现力的重要性,在音乐讨论中,音乐的时间维度却往往被忽视。也就是说,即使知道音乐这一艺术形式必然在时间中存在,在音乐讨论中我们依旧往往疏忽时间。正因如此,我们需要绕道物理学,借用"时间箭头"这一并非原生出自音乐领域的术语。哲学家奥古斯丁对时间的描述放在音乐领域的上下文中尤其应景,精准捕获(capture)了时间是如何无处不在,却又同时在严肃的讨论中被忽视:

时间是什么?我们显然在提到时间的时候知道它是什么。我们在听到人们说时间的时候也知道它是什么。所以时间到底是什么?如果没人问我,我知道。但如果让我向他人解释,那么我一无所知。④

音乐理论中有一个隐含的潜规则:音高(pitch)比节奏(rhythm)更重要。因此,相对于音高来说,人们对节奏等时间上的参数的理论研究较少。这一现象几乎贯穿了整个西方音乐理论的发展史。我们可以将乐理的发展以调性音乐作为分界线,粗略分为三个时期:

·前调性(pre-tonal)时期:大致覆盖了文艺复兴到巴洛克的这一段时间,代表为各种对位理论。

·调性(tonal)时期:大致覆盖了巴洛克到浪漫主义的这一段时间,代表为各种调性(tonal)和声/功能(functional)和声理论。

·后调性(post-tonal)时期:大致覆盖了浪漫主义之后的时间,代表为各种无调性(atonal),泛调性(pan-tonal),十二音技法(twelve-tone technique),序列主义(serialism),音乐集合论(music set theory),

新里曼理论(Neo-Riemannian theory)等等。

我们发现,上述的代表理论中,对音高(而非节奏)的研究占据了主导:对位法是基于协和音程和不协和音程的概念发展出来的理论,核心就是如何控制和解决不协和的音程;调性和声,顾名思义,核心概念是调性,也就是说不同的和声如何在调性这个画板上交互;十二音技法所依赖的音列(tone row)以及对应的操作即 transposition/inversion/retrograde/retrograde-inversion,都是主要基于音高(pitch-centric)的概念。固然在这些理论中,我们可以找到对节奏的论述,例如在分类对位法(species counterpoint)中,对位的规则需要考虑节奏的强拍弱拍。然而不置可否的是,这些理论都主要围绕着音高概念展开,节奏往往是次要(secondary)概念。

除此之外,涉及到节奏的理论,往往是基于音高的理论的衍生(derivative),或者说延伸(extension)。也就是说,一个理论往往先基于音高概念来打造,构造出基于音高的符号表达(symbolic representation)和符号操作规则(rules for symbolic manipulation)。之后,我们试图将这一基于音高的理论直接移植(port)到节奏参数上,用操作音高参数的规则来操作节奏参数。在 20 世纪中期兴起的"完全序列主义"(total serialism)就是一个例子:其主张就是将序列主义对 12 音的操作,应用到音高之外的参数,例如时值或是力度。我们可能会说,完全序列主义的兴起不就是对音符时值,乃至节奏的关注了?不就是意识到了单纯基于音高的序列主义的局限性,而作出的改良?恰恰相反,这刚好体现了对时间的不重视:完全序列主义,可以被看作是序列主义,或者说十二音体系的进一步发展,进一步延伸。我们可以试作一个解读:音高更加重要,所以要优先专门基于音高参数发展量身定做的理论;节奏更次要(subordinate),所以不需要专门基于节奏参数发展理论,而是直接移植基于音高的序列主义即可。正如序列主义中的关键概念"音列"的字面意思所示,序列主义在发展之初关注的只是音乐的音高结构,并没有考虑时间结构。因此,生硬地将对"音列"的操作移植到非音高的参数上,必然会忽视音乐的时间

④Saint Augustine, *The Confessions*, trans. Henry Chadwick, *oxford worlds classics*, Oxford University Press, 2009, p. 230.

结构，因为理论在设计之初就没有关注时间结构。例如，音列的概念在设计之初，是基于音高的符号表达：音列中的元素是音高集合（pitch class）。那么，当音列被用来指代非音高参数，例如时值的时候，就会产生很多对合理性（justification）的质疑。例如，大小为12的音列，用在音高方面的时候合理，因为数字12对应着一个八度包含的12个音高集合：大小为12的音列刚好可以填满（saturate）所有的音高集合。但用大小为12的音列来指代时值，则更像是一个随意的选择（arbitrary choice），缺乏合理性。

将对节奏的研究看作是对音高研究的一个副产品，我们就不难理解另一个理论的做法。在新里曼理论的著作《广义音程与变换》（*Generalized Musical Intervals and Transformations*）中，作者大卫·列文（David Lewin）写道：

> 本章的出发点是一个图例，包含了在音乐符号空间中两个"点"：s和t。我们用一个被标为 i 的箭头来表示从 s 到 t 的测量，距离，或是运动。这一箭头直观地描述了在多种音乐空间中的情况，而当 s 和 t 是音高集合的时候，i 也被称作"从 s 到 t 的音程"。[5]

这一著作提出了一个概念：广义音程系统（Generalized Interval System, GIS）。顾名思义，GIS 基于音程（interval）这一概念打造，并推广（generalize）到音高之外的参数，例如节奏。在音乐讨论的英语语境中，interval 一词有着无歧义的指代：它优先指代基于音高的音程（pitch interval）。也就是说，在缺省情况下，即不加任何限定语（qualifier）的时候，interval 一词指代音程关系。如果我们想要表达时间维度的距离关系（temporal interval），那么需要加上限定词前缀，例如"time-span" interval（即"时间跨度"层面的距离）。或许是作者也意识到了我们对音高的重视是如何反映到 interval 一词上的，专著呈现 GIS 模型的顺序非常直观：以音程为出发点定义 GIS 模型（因为 interval 一词优先指代音程），然后再将这一模型推广到节奏参数。

申克分析（Schenkerian analysis）也可以让我

们一窥"音高比节奏重要"这一想法是如何体现在申克分析的理论思维中。申克分析中的一个重要概念是申克图（Schenkerian graph）：一种简化的谱面，被用来描述曲目的深层和声及曲式结构。通过观察，我们可以发现一个很有意思的现象：申克图透露出了更多有关音高和声关系（例如深层结构（Ursatz）这一概念），而不是有关时值的结构。申克分析可以被看作是一个简化（reduction）过程：从前景，即可见的谱面，层层剥茧抽丝，我们逐渐失去了音符的时值信息，最终剩下的就是作为骨架的和声，以及高音部的旋律。也就是说，如果我们将申克分析的这个简化过程看作一个动态的过滤过程，那么大浪淘沙，音符的时值信息，或者说时间信息在这个过程中被过滤掉了。我们因此可以尝试解读：在申克分析中，相对于基于音高的和声和旋律，基于时间的时值更加不重要，因为不重要，所以可以省略（omit）掉。用申克分析的话来说，音乐的节奏结构可以被看作是前景，是流于表面的修饰，因为不具有结构上的重要性，所以省略。我们甚至可以推测，正是因为时值在申克分析中不重要，所以申克图中音符的时值更像是滥用符号（abuse of notation）行为，并不对应音乐的时间结构。正如菲利克斯·萨尔彻（Felix Salzer）在申克分析理论专著《结构化聆听》（*Structural Hearing*）的脚注中提示：

> 在申克图中用二分音符来标注和弦并不表示时值，而是用来区分和弦的不同功能。[6]

一部分学术界的人也注意到了这一乐理中忽视时间的现象。比如，1985年春季版的期刊《乐理光谱》（*Music Theory Spectrum*）第七卷将一整期用于讨论"音乐里的时间和节奏"。编者在引言中就开篇说道，音乐的时间维度很重要，但同时又是相对来说学术研究较少触及的领域[7]。在同一卷的第72页，作者直言不讳道：音乐方面的时间并没有被广泛认可为一个单独的研究领域。他之后直接如数家珍般地给出了证据：

《新格罗夫音乐与音乐家辞典》没有"时间"这一词条；《国际音乐文献资料大全》（*RILM*）没有时间这一单独的分类；《音乐索引》最近才开始在"时

间"这一词条下加入内容。⑧

除此之外，在 1960 年的专著《音乐的节奏结构》(*The Rhythmic Structure of Music*) 中，作者们毫不委婉，直接在序言中激进提到：

> 在音乐家的训练中，音乐的时间维度从文艺复兴时期以来就几乎被完全忽视。有关和声和对位的教科书极多，然而有关节奏的教科书则完全没有。⑨

或许我们会说，这种忽视可能只是一个巧合。比如，可能是因为时代局限性，或者西方古典音乐的文化局限性，才让人类文明的某一特定的时空子集忽视音乐中的时间。但即使如此，也足以证明问题：不论是因为时代的局限性也好，西方古典音乐文化的缺陷 (bug) 或者说特性 (feature) 也罢，一个本质上基于时间的艺术形态，居然在研究中会忽视时间，本身就匪夷所思。

在音乐讨论中对时间的忽视，除了匪夷所思，令人兴奋：它给我们的音乐讨论指出了一个潜在有前景的方向。与其将对时间的忽视看作一个缺陷，或许更为恰当的是将它看作一个机遇。一方面，时间维度对音乐至关重要；另一方面，音乐讨论中往往缺失了时间维度 (因此，我们有必要将目光投向其他对时间维度研究较多的领域，来进行参考和借鉴)。这种客观重要与主观忽视并存的状态，给潜在有价值 (fruitful) 的研究提供了机会，并直接成为了驱动本文撰写的主要动机。

## 哲学和科学中的时间箭头

上文提到，时间的特殊性质在于它的指向性，即时间箭头。我们可以进一步将这一性质分解 (decompose) 为两个命题来描述：

1. 可区分性条件 (distinguishability condition)："过去"和"未来"有着本质上的不同。

2. 唯一指向性条件 (unique orientability condition)：时间的流逝存在着唯一的一个固定的方向，即从"过去"指向"未来"。这个固定的方向完全先验 (a priori)，不以人的意志为转移。

这一分解的优势在于，在后文我们可以看到，上述两个命题很好地对应了音乐中常见的现象，为音乐分析提供了切实的指导。然而，要更好地理解上述两个命题，我们需要先跳出音乐，甚至跳出艺术，透过哲学和科学这两个对时间箭头讨论较多的领域，来进行参考和借鉴。

### 一、哲学中认识论对可区分性条件的解读

我们如何区分 (distinguish) 过去和未来？认识论围绕着"知识是什么"这个问题展开，我们可以对"区分过去和未来"这一问题作出适当解答。从认识论的视角解读，过去和未来的区别在于"对过去"和"对未来"的知识的讨论：我们如何了解过去，获取对过去的知识；以及我们如何了解未来，获取对未来的知识。在哲学专著《时间与偶然性》(*Time and Chance*) 中，作者大卫·阿尔伯特 (David Albert) 进一步提出：与其说过去和未来有着本质的不同，更合适的说法是，我们了解过去和了解未来的认知过程和方法有着本质的不同⑩。也就是说，过去和未来的区别，只是对于求知这一动态过程的不同所带来的产物。

我们记得过去，却不记得未来。这一看似不言自明的常识有着另一种描述：我们主要依赖"记录" (record) 来了解过去。也就是说，对于过去，我们有一种叫作"记录"的东西，记载着过去的事件；然而对于未来，则无存在"记录未来"的可能性。我们并非记得 (recall) 未来，而是预测 (predict) 未来。记录和预测是本质上不同的两种行为，并非同一行为面向过去和未来。阿尔伯特指出：

> 所有我们知晓的未来 (或是更广义来说，能够知晓的未来)，都在原则上依赖"预测"这一行为。某些 (并非所有，甚至可以说只有很少一部分) 我们知晓的过去 (例如行星过去的位置) 也类似于知晓未来，依赖"回溯"这一行为。然而，绝大部分我们知晓的过去都来自于"记录"。⑪

也就是说，我们一共有三种了解过去和未来的方式：读取记录 (record-reading)(即记忆)，根据现状倒推回溯 (retrodiction)，以及预测 (predic-

tion)。我们有两种了解过去的方式,却只有一种了解未来的方式。回溯和预测是对应的概念(即同一行为,面向不同的时间点),而读取记录则没有相应的对应概念。同时,因为读取记录远比倒推回溯要高效,我们将读取记录当作主要的了解过去的方式,基本不采用回溯的方法。因此,过去和未来的最大区别就在于"记录"这一个特殊的、没有面向未来的对应概念的认知过程。

通过读取记录了解过去的高效体现在,它极大节省了我们了解过去所需的计算成本。阿尔伯特提出[12]一个思维实验(thought experiment)用以举例:假设我们想知道台球桌上的五号球在最近十秒钟有没有碰到其他的球。如果我们通过回溯的方法来了解,那么所需要做的工作跟预测五号球在未来十秒钟会不会碰到其他的球是一样的(因为微观层面的物理规则具有时间反演对称性);我们需要考虑桌上所有台球的状态,并且根据运动方程(equations of motion)进行复杂的演算。然而,如果有记录,那么我们只需要读取一个大小为一位元(bit)的信息即可知道答案:五号球在最近十秒钟是否有碰到其他的球,1为是(true),0为否(false)。

记录的存在,隐含着一个重要的假设:我们相信记录能够可靠地给我们有关过去的知识。具体来说,我们相信过去的事件能够在某个局部空间留下痕迹(trace),让某些物件成为"记录的载体"(record-bearing device),以至于我们可以仅仅通过观察这一局部空间来作对过去的推断(inference)。例如,当我们看到指纹痕迹,那么我们几乎可以肯定,指纹痕迹的来源是且只能是在过去的某一时刻,有这一指纹的人留下了指纹痕迹,而几乎绝无可能是风吹日晒所自发形成的。有学者提出过类似的例子[13]:沙滩上的脚印,告诉我们有人曾走过,因为脚印这一记录的存在来自于"过去有人走过"这一事件。然而,一片平整的沙滩却无法告诉我们未来这片沙滩是否将会有脚印,原因很简单:对于过去,脚印的存在只对应"有人走过"这一种可能的事件;然而对于未来,平整的沙

滩对应的是无限种可能的事件,例如有脚印,有沙堡,或者依旧平整。

上述的例子隐约点出了我们为何信任对过去的记录,而不信任(甚至无法想象)对未来的记录:一个记录,往往只对应过去的一种可能性(也就是这个记录所记载的事件);然而,未来的可能性更多,所以很难一一对应某个纪录。也就是说,过去和未来的区别体现在:相比于过去,未来的可能性更多,更加无序。在这里,我们可以观察到热力学中往往提及的概念:熵。实际上,隐约感受到的热力学与时间关联并非空穴来风,阿尔伯特指出:

事实上,热力学第二定律之所以成立的原因跟我们能够通过预测/回溯之外的方法了解过去的原因是同一个。之所以存在对过去而不是未来的记录,正是因为我们的经验似乎在肯定一个"过往假说",但没有任何"未来假说"。[14]

也就是说,认识论时间箭头跟热力学时间箭头同源,是一体两面。热力学第二定律的有效,以及"记录过去"的可能性,是源自同一个原因:我们对宇宙过去的状态存在一种特殊的假设,阿尔伯特将其称作"过往假说"(past-hypothesis)。这一假设在热力学和认识论中有着不同表述:在认识论中,是上文所提到的假设(即,我们相信记录能够可靠地给我们有关过去的知识);在热力学中,则是假设宇宙的宏观初始条件是低熵状态。

二、科学中热力学对唯一指向性条件的解读

从词源(etymology)角度,热力学是"时间箭头"一词的最初来源。也就是说,时间箭头在热力学中是真正的原生概念。在著作《物理世界的本质》(The Nature of the Physical World)中,物理学家爱丁顿用箭头来比喻时间:

我们先随便画一个箭头。如果我们沿着箭头的方向前进,并且发现世界的状态中,随机无序的元素越来越多,那么我们将箭头指向的方向称作"未来";如果发现随机无序的元素越来越少,那么我们将箭头指向的方向称作"过去"……我用"时间箭头"来描述这一时间的单向性质,这一性质在空间中没有相应的对应。[15]

我们往往会用"熵"这个概念和热力学第二定律来解释时间箭头。热力学第二定律中的核心概念，也就是熵，可以被看作是对一个系统的"无序度"的衡量：熵越高，就越混乱，越无序。而热力学第二定律被用来描述熵如何随着时间而变化。这一定律的表述方式有很多，其中一种是：

一个孤立系统中的熵只会随着时间增加或者不变，但绝不可能减少。[16]

也就是说，在没有跟外界交流的情况下，一个孤立系统倾向于向着越来越混乱越来越无序的状态演化，而这个倾向规定义着时间的指向性，即时间箭头。在日常生活中，我们可以感受到热力学第二定律的效果，并诗意地将其描述为世界的恶意倾向。不论是我们在生活中常说的"人生不如意事十之八九"，或者是音乐家对上台演出的敬畏，都体现出了我们普遍存在的对世界恶意倾向的感知。正因为事态倾向于向恶化方向发展，我们的常识告诉我们，好事多磨，让事态进步比退步更难：逆水行舟，进步需要耗费精力心血，而退步则只需要不作为。

然而，我们需要避免过度诗意：世界的恶意倾向，并不是指我们需要拟人化世界，仿佛世界在有目的的意志里跟我们的愿望对着干。之所以感受到世界的恶意倾向，是因为我们只珍视（appreciate）所有的可能的极小一部分，并将其称作"有序"（orderly），所以即使世界在无差别做选择，我们也会感觉事与愿违。也就是说，世界的恶意倾向是一个比例问题：有序的可能远比无序的可能要稀少。亚里士多德在《尼各马可伦理学》（The Nicomachean Ethics）写道：

失败的途径很多（因为邪恶隶属于"无限"的类别，正如毕达哥拉斯派的人所提出，而正义隶属于"有限"的类别），而成功只有一种途径（这也是为什么前者简单，后者困难，因为偏离准心容易，正中准心困难）……人只有一种向好的途径，却有多种向坏的途径。[17]

同理，在小说《安娜·卡列尼娜》（Anna Karenina）中，托尔斯泰开篇就写道：

幸福的家庭都是相似的；不幸的家庭各有各的不幸。[18]

热力学中的宏观态（macrostate）和微观态（microstate）帮助我们更好地理解世界的恶意倾向，并为其提供了一个量化的描述：有序的可能远比无序的可能要稀少。微观态是一个对系统具体的状态的完整描述（complete specification）。微观态往往涉及到对大量细节参数的详尽描述，例如记载一个容器内所有的分子的位置（position）和速度（velocity）。然而，对于要解决的问题，我们可能不想（disinterested）也不能（infeasible）知道一个系统的微观态：例如知道有分子的位置和速度需要耗费大量的计算力。也就是说，很多不同的微观态对于面对的具体问题来说等价（equivalent），且不可区分（indistinguishable）。因此，顾名思义，我们往往可以用宏观态来描述系统的宏观状态：一个宏观态往往对应多个等价的微观态。从数学角度来说，宏观态就是微观态的等价类（equivalence class）：宏观态定义了一个微观态集合中的等价关系（equivalence relation），为划分（partition）微观态集合为一个个的宏观态等价类。

接下来，我们关注的是宏观态的大小（cardinality），也就是说一个宏观态对应/包含多少个微观态。宏观态的大小量化了熵这一概念：越大的宏观态，对应着更大的熵。物理学家玻尔兹曼给出了熵的另一描述：

熵衡量的是从宏观角度来看无法区分的微观排列组合的数量。[19]

用宏观态描述熵，进一步解释了我们的直觉，即世界具有恶意倾向，事态倾向于向毁灭演化：对应"好事"/"有序"的宏观态是一个相对较小的等价集，所以即使世界以相等的概率选择微观态，落在"好事"/"有序"宏观态的几率也相对较小。

正如前文中阿尔伯特指出，热力学时间箭头和认识论时间箭头因为所共同假设的过往假说而具有同源关系。所以从某种角度来说，我们之所以能够欣赏在时间中流动起伏的音乐，其可能性就恰恰来自看上去和音乐无关的热力学时间箭头。

## 音乐方面的应用与启示

从音乐的角度,将上文提到的可区分性条件和唯一指向性条件分别展开,我们可以进一步得到两个命题:

1. 可区分性条件的音乐角度展开:音乐事件的功能和效果取决于它在曲目的剧情发展中所处的时间上下文。

2. 唯一指向性条件的音乐角度展开:时间箭头决定了曲目中戏剧高潮的放置位置。

一、可区分性条件——复现

要理解可区分性条件的音乐角度展开,一个很好的切入点就是控制变量:观察同一音乐事件,如何根据其所处的不同时间上下文,而呈现出不同的性质。同一音乐事件在不同时间点上的出现,就是音乐里我们耳熟能详的概念:重复(repetition),或更加诗意地说,复现(recurrence)。

此文以巴赫的宗教作品为案例,并假设巴赫的宗教作品往往具有"音乐外的指代"(extramusical reference):音乐事件往往对应音乐外的某些含义。具体来说,音乐外的指代,让我们可以更好理解"复现"这一音乐概念,进而理解音乐中的时间箭头。巴赫的宗教作品,让我们可以提出一个更加具体的对应:同一音乐事件在不同时间点上的出现,如果代入到"生命的从无到有"这一过程在不同时间点上的出现,那么则可以自然得出巴赫的宗教思维中的两个重要概念,即"诞生"(birth)和"重生"(resurrection)。诞生和重生,描述的其实是同一个过程,即生命的从无到有:如果这一过程发生在生命的无奈和苦难之前,是诞生;如果在之后,则是重生。

在基督教语境下,耶稣的生命轨迹由三个重要时刻调控:诞生、死亡、重生。这三个时刻体现在了基督教历法(liturgical calendar)上,对应着三个重要节日:圣诞节(Christmas),受难日(Good Friday),复活节(Easter)。已知巴赫的宗教情结,我们可以在不过多担心过度解读(intentional fallacy)的情况下适当猜测:在巴赫的作品中,存在

对"诞生/死亡/重生"的特意关注,并在音乐层面有着可观测的体现。此外,我们可以猜测,巴赫为重要宗教节日所作的曲子承载着他更多的心血和精力。为此,《马太受难曲》这一致敬受难日的作品,就自然成为了一个很好的分析切入点。

在巴赫的宗教作品中,时常穿插出现的众赞歌(chorale)具有特殊的意义。首先,在听感方面,众赞歌能够在所处的前后文中脱颖而出,快速抓住观众的注意力;织体(texture)方面,众赞歌的齐唱(homophonic)具有鲜明的节奏(所有声部节奏统一)及复声(所有声部相同程度的参与),得到突出的音量)特征;内容方面,众赞歌往往采用观众耳熟能详的旋律,甚至传统习俗会鼓励观众也加入众赞歌段落的合唱(路德宗相信,通过聚在一起高歌,人们可以通过音乐更好地参与祷告[20],建立起人与人,与上天的联系,达到类似于天人合一的境界)。上述音乐特征,让观众可以迅速将众赞歌段落从前后文中分离,并视作一个特殊段落。其次,更为重要的是,穿插出现的众赞歌在巴赫的作品中,往往起着评述(commentary)的作用:它代表着群众(congregation)的声音,暂时跳出/暂停了作品的剧情(narrative)发展,取而代之的是描述观众对剧情的围观和感悟。用今日的比喻来说,众赞歌就像是电视节目中偶尔给到台下观众的镜头:观众的喜怒哀乐,作为对台上表演的评述,也成为了节目的一部分。正如大卫·希尔(David Hill)观察到:

一般来说,众赞歌被认为是代表了巴赫设想出的观众所应有的感受……巴赫将剧情植入到了观众的脑中,许多评论家也意识到了这一点,并且认为众赞歌描述了观众的想法。[21]

对于听众,众赞歌往往具有指导作用:它是作曲家所给出的"官方提示",描述了观众"应有"的感触。此外,众赞歌的时刻往往对应全曲中的重要时刻,因为只有重要才值得驻足品味。因此,我们也可以通过巴赫作品中的众赞歌来高效把握全曲的整体结构。具体来说,我们可以将全曲的众赞歌段落提取出来,并将其看作自成一套的组曲(suite),那么我们则得到了一个描述整体结构的大纲(outline),一个音乐分析角度的简化(analyti-

cal reduction)。

以众赞歌为线索，就能凸显出《马太受难曲》的一个特点：基于同一个旋律的众赞歌，在全曲中穿插出现了整整五次。被称作"啊，受伤流血尽受嘲弄的头颅"(O Haupt voll Blut und Wunden)的旋律以众赞歌的形式出现在此曲中的第 15（见谱例 1）、17（见谱例 2）、44（见谱例 3）、54（见谱例 4）、62（见谱例 5）乐章。这一旋律的出现，也对应着剧情中的几个重要时刻：耶稣在橄榄山(Mount of Olives)（旋律出现两次），审判，受难，死后。一个众赞歌在同一作品中频繁重复，在巴赫的作品中

罕见且不寻常(anomaly)。单单是这一旋律重复的次数就能让观众记忆深刻，这也是音乐中的"重复"最为简单但是强大的功能：重复作为对记忆的强化(reinforcement)。同时，希尔指出，这一旋律贯穿全曲的重复出现，会让听众不自觉将这一重复出现视作一个系列(series)：

　　[这一重复出现的众赞歌旋律]是马太受难曲中最显著的元素之一，也是大多数听众能够直观感受到一个"系列化"的旋律。作为一个系列，这一贯穿全曲的旋律指引着众人的思绪，并在剧情最为悲剧的一刻达到高潮。[22]

谱例 1　巴赫《马太受难曲》，BWV 244，第 15 乐章[23]

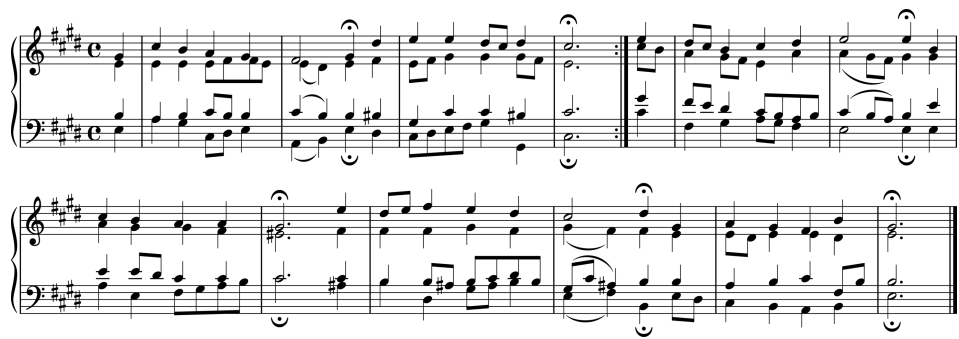

谱例 2　巴赫《马太受难曲》BWV 244，第 17 乐章[24]

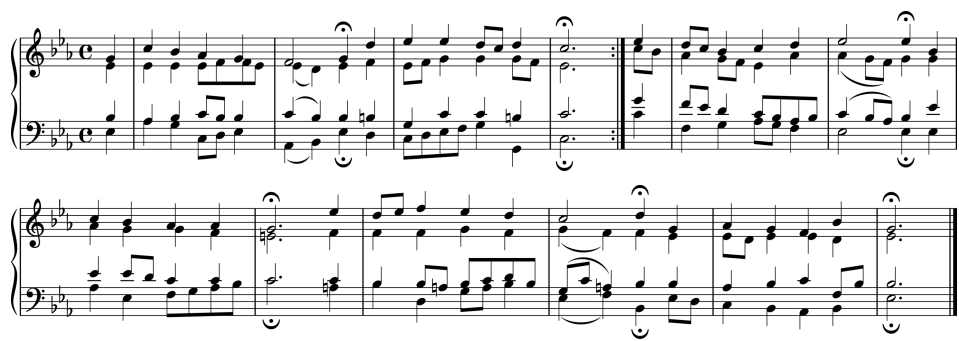

谱例 3　巴赫《马太受难曲》BWV 244，第 44 乐章[25]

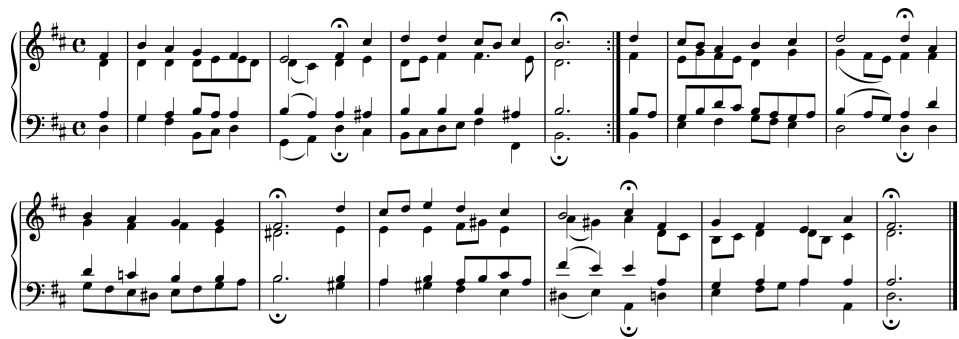

谱例4 巴赫《马太受难曲》BWV 244,第54乐章㉕

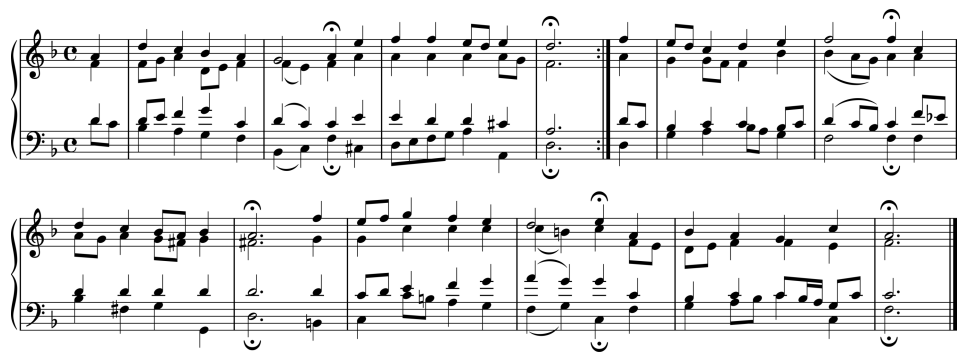

谱例5 巴赫《马太受难曲》BWV 244,第62乐章㉖

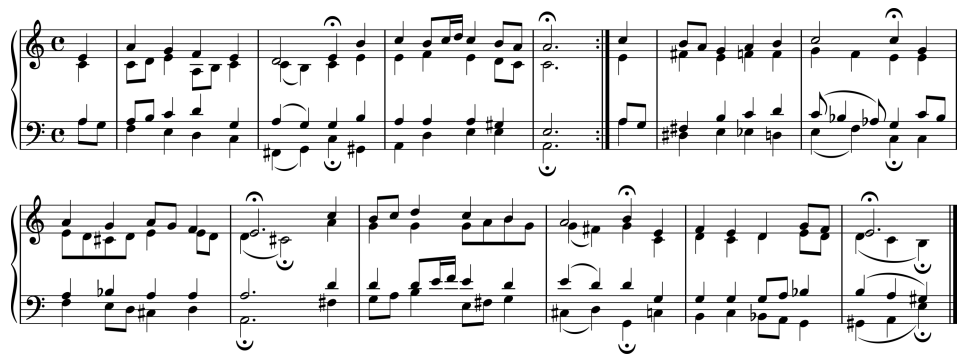

结合前文所说,我们可以把这一旋律的五次出现当作全曲的大纲,一个剧情发展的缩影:同一旋律的重复出现描述了观众的心境如何随着剧情发展而波动。《马太受难曲》是一个特例,特殊在这一大纲由相同音乐素材的复现组成。复现起到了控制变量的效果:通过比对这一旋律五次出现的异同,我们可以更好地理解观众心境的波动,进而理解全曲的音乐结构。更具体来说,我们尤其关注众赞歌在五次复现中,通过重复所建立起来的音乐规律(regularity/pattern)被打破的地方,因为在音乐感知中,我们对打破已有规律的段落尤其敏感。

在五次复现中,最后一次显得尤其特殊。旋律方面,前四次一样的旋律,在第五次中有了不一样的收尾(见谱例6)。不论是新增的十六分音符,还是更加婉转(inflected)的旋律走向,都给旋律增添了一种流动和不安。和声方面,第五次也更加不安:引入的半音(chromaticism)(见图7)可以被看作是对之前四次中和声的扭曲和变形。用"扭曲/变形"等主观词汇描述这里的和声并非过度解读:巴赫承袭文艺复兴时期的绘词法(word-

painting)传统,往往会用半音来描述内心的扭曲和挣扎。"扭曲"一词很恰当描绘了这一段落对应的剧情中的情绪:在耶稣死后,众人心如刀绞,所感受到的弱小,可怜,又无助。

谱例6 巴赫《马太受难曲》BWV244㉗

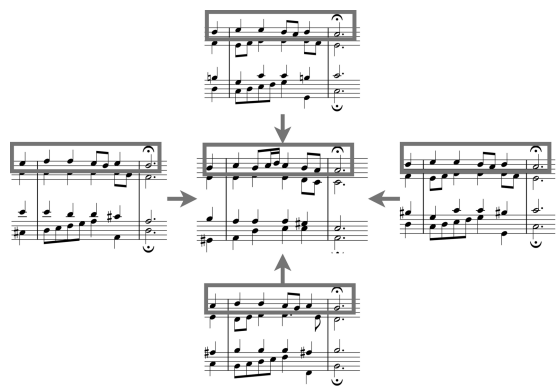

我们可以试解读众赞歌在第五次出现时通过旋律及和声打破建立起的规律背后的含义。五次复现所勾勒出的音乐发展轨迹,暴露了此曲的一个"问题"(这一词在这里不指缺陷,而是主动提出引人深思的问题):旋律跟和声的不安,用技术性描述来说则是不稳定(destabilizing),让我们觉得

谱例 7　巴赫《马太受难曲》BWV244，第 62 乐章[29]

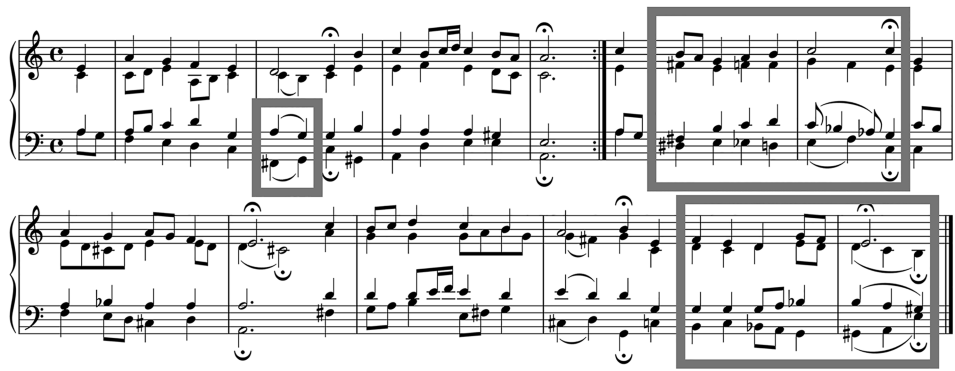

《马太受难曲》并不完整，而是有着一个开放式结尾。此外，它的不完整并非缺陷，而更像是一个有意为之的设计。这一设计，就像是电影结尾放置的彩蛋，透过重新燃起的悬念，预示着续集的存在。结合巴赫的神学观念，我们不难看出，他通过《马太受难曲》的开放式结尾，来给听众以希望：死亡不是生命的终结。在基督教语境下，耶稣将在复活节那一天重生，因此《马太受难曲》的叙事并未结束。

**可区分性条件的案例：对《圣诞节清唱剧》做重新定位**

那么，《马太受难曲》的续集在哪里？从本文所讨论的时间箭头出发，带着对时间上下文的敏感，我们试作一个反直觉的解读：《圣诞节清唱剧》（BWV 248）是《马太受难曲》的延续。这一解读的反直觉之处在于，根据日历时间线，圣诞节先于受

难日。将献礼圣诞节的曲目看作致敬受难日的曲目的延续，违背了日历角度的先后顺序。然而从音乐角度，以众赞歌为线索，《圣诞节清唱剧》可以被看作是《马太受难曲》的延续和补全，为开放式结尾提供了一个圆满的结局。

《马太受难曲》中重复五次的众赞歌旋律，在《圣诞节清唱剧》中再次出现两次：第 5（见谱例 8）和第 64 乐章（也是最后一个乐章）（见谱例 9），分别对应着全曲的第一个和最后一个众赞歌。单单是首尾呼应的放置，就能凸显出这一旋律对于此曲的重要性：正如 Alfred Durr 所指出，这一首尾呼应，可以被看作是巴赫在作曲中达成主题统一性（thematic unification）的手段[30]。在认识到了这一旋律分别对于此曲和《马太受难曲》的重要性之后，我们可以将两首作品当作一个整体叙事，考虑总共出现七次的众赞歌旋律。

谱例 8　巴赫《圣诞节清唱剧》，BWV248，第 5 乐章

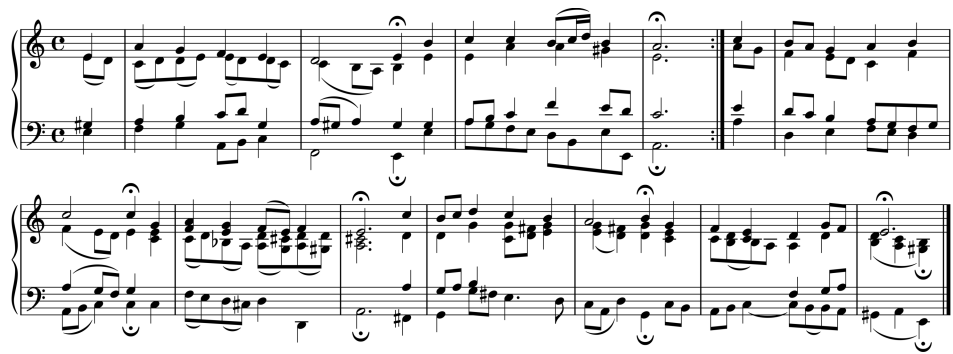

尤其值得关注和比对的是这一旋律在《马太受难曲》中的最后一次出现和在《圣诞节清唱剧》中的第一次出现：两者的相似，让我们感受到了叙事上的重叠，仿佛《圣诞节清唱剧》捡起了

前曲未讲完的故事，继续娓娓道来。这让我们有足够的信心猜测，音乐方面，《圣诞节清唱剧》可以被视作《马太受难曲》的延续：旋律方面，两者都采用了特征显著的十六分音符（见谱例 10）；和声

谱例9 巴赫《圣诞节清唱剧》，BWV248，第64乐章①

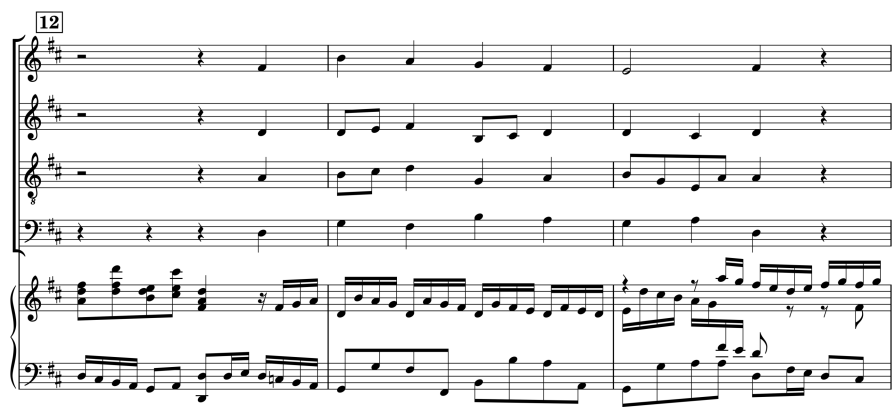

谱例10 《马太受难曲》与《圣诞节清唱剧》比对乐章的第3小节②

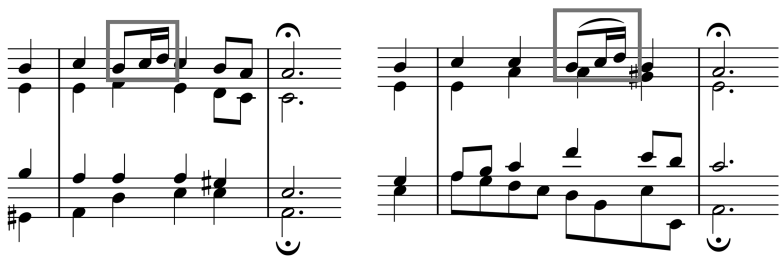

方面，两者结尾的终止式采用了完全一样的和声进行（见谱例11），即变格终止（plagal cadence），在听感上极具辨识度。我们甚至可以说，两者实际上是一模一样的重复：这一旋律在《圣诞节清唱剧》中第一个众赞歌的出现，起到的就是前情提要，或者说类似于"上回书说到"的作用。上回书说到哪里？说到了《马太受难曲》中那悬而未决的变格中止。

谱例11 《马太受难曲》与《圣诞节清唱剧》
比对乐章结尾的终止式③

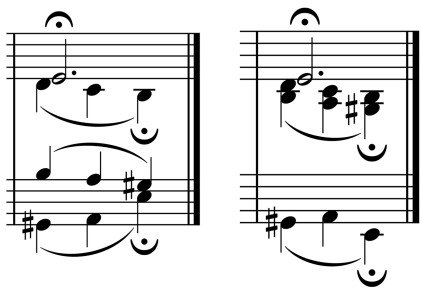

同时，《圣诞节清唱剧》的结尾补全了前曲的开放式结尾：之前的不安和悬念烟消云散，取而代之的是圆满的尘埃落定（finality）。在众赞歌旋律的七次出现中，《圣诞节清唱剧》的结尾特殊在，乐队的身份不再是伴奏，而是跟合唱团交织，形成对话交流。乐队的加入，把全曲的情绪推到了一个圆满喜庆的高潮：喜悦的气氛惊天地泣鬼神，以至于乐队所代表的花花草草都加入了欢庆。我们因此可以把这一乐章看作是一个众赞歌旋律经过七次重复迭代之后，演化出的完全展开形态，如同孔雀开屏，大鹏展翅，牡丹盛开。

因此，对时间箭头的讨论让我们对时间上下文敏感，并可以作出一个音乐分析上的解读：同一音乐事件（在这里，即是《圣诞节清唱剧》这一作品），根据其所处的不同时间上下文（即，先于或者后于《马太受难曲》），而呈现出不同的性质（即象征"诞生"或者是"重生"）。也就是说，《圣诞节清唱剧》因为时间箭头的缘故，具有双重身份：

1. 前置（放置在《马太受难曲》之前）：当它被看作一个独立的作品，那么它（顾名思义）象征着圣诞节，即耶稣的诞生。

2. 后置：当它被看作《马太受难曲》的延续，那么它象征着耶稣的重生。

这一结论，从音乐的角度陈述了一个神学命题：诞生和重生是一体两面。

① 本谱例为巴赫《圣诞节清唱剧》选段，在乐队部分的前奏后，合唱团加入的部分。
② 谱例左为《马太受难曲》第62乐章，右为《圣诞节清唱剧》第5乐章。谱例的方框为笔者批注，点出旋律特征。
③ 谱例11，其左为《马太受难曲》第62乐章，其右为《圣诞节清唱剧》第5乐章终止式。

## 二、唯一指向性条件——高点的后移

当我们用箭头来比喻时间，我们往往会假设它具有某种"推动力"：箭头顺着它所指向的方向把物体往前推。也就是说，在直觉上，箭头除了描述方向，还往往描述着某种力量。在数学/物理中，这一直觉由力场（force field），以及更广义的向量场（vector field）捕获：我们用箭头来描述力量，而这里的箭头不再是一个比喻，而是表示向量（vector）这一精确的概念。向量，顾名思义，具有方向（direction）和量级（magnitude）。我们之所以感受到箭头具有推动力，就是因为它捕获了方向和量这两个力的重要性质。

时间箭头推动着什么？在音乐中，与其泛泛而谈提及例如"时间箭头推动着音乐的戏剧张力"，我们可以关注一个原生于音乐的概念：一首曲子中的戏剧高点，并尤其关注它在曲目中的放置位置。我们甚至可以说，对一首曲目的戏剧高点的把控，直接决定了全曲的音乐表现力。演奏者因为直接负责音乐表现力，往往对一首曲目的戏剧高点尤其敏感。例如，根据记载㉞，拉赫玛尼诺夫尤其强调一首曲目中的"巅峰时刻"（culminating point），演奏者务必确保这一时刻的精彩，因为错失这一刻就会毁掉整场演出。

因此，我们可以提出一个既诗意又具体的主张：时间箭头推动着一首曲目戏剧高点的放置位置，带来了高点的后移（delay）（见图1）。也就是

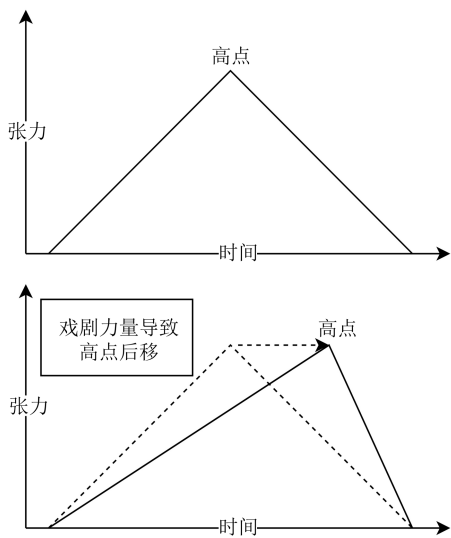

**图1　高点的后移带来的时间不对称性，直观体现为左右不对称的图形**

说，时间箭头（在音乐中）顺着它所指向的方向（即未来）把物体（即戏剧高点）往前推。对此，我们关注的是一个对这一动态过程完整的描述（complete description）：之所以被称作后移，是因为我们认为高点原本应当落在曲子当中的某处，但受到某种被称作"时间箭头"的神秘戏剧力量影响，被逐渐推移到了之后的某处。这一动态过程的结果，即后移过的高点，也给了我们一个最为直观的对"时间不对称"的描述（也是本文撰写的原始动机之一）：曲子中段落比例的不对等，因为高点的后移导致高点前的前半段比高点后的后半段要更加冗长，体现为图1所呈现出的视觉方面的左右不对称。

高点的后移作为一个过程，也可以被看作是一个表象，是曲式层面的生长扩张过程所呈现出来的一个症状。具体来说，我们可以将观测到的曲式结构（formal structure）看作是一个动态生长过程的产物。其中，高点的后移无非是表明，这一动态生长过程优先（prioritize）关注高点前（pre-climax）的部分，并因为高点前部分的优先生长，将高点往曲末方向挤压。宇宙的膨胀是一个尤其贴切的比喻：高点的后移，就像是星星的远去，所揭露的是一个扩张膨胀的过程。

值得注意的是，将曲式结构看作动态生长过程的产物，是一个尤其有机主义的视角。这一视角在申克分析的思想中已有端倪。我们在谈及申克分析的时候，往往强调的是申克分析作为一个简化过程（reduction），可以被看作是一个自下而上（bottom-up）的描述：从前景（也就是充斥着具体细节的下层）出发，层层筛选，得到曲目的背景（最为本质简洁的上层结构），或者用申克分析的专有术语来说，得到曲目的深层结构（Ursatz）。然而，我们也可以掉转思路（mindset），获得一个自上而下（top-down）的描述：从背景的深层结构，层层展开（unfold），动态生长，得到缤纷多彩的前景，也就是我们所感知到的音乐本身。这一自上而下的描述，让申克分析成为了一个生成过程（generation），比自下而上的描述更加有机主义，甚至更符合申克分析本意。例如，申克分析中的两个重要概念，展开（elaboration）和延长（prolongation），都

---

㉞David Dubal, *The Art of the Piano: Its Performers, Literature, and Recordings*, Hal Leonard Corporation, 2004, p. 286.

是描述如何对背景中的结构进行修饰展开,填充细节,获得下一层前景中的元素。这两个概念隐含了背景是前景的先决条件(prerequisite)的依赖关系(即,前景是对什么的 elaboration/prolongation),几乎规定了我们必须按照从背景向前景推进的顺序来诠释申克分析,否则根本无法定义这两个概念。

　　沿用申克分析的自下而上描述,我们从深层结构出发,这一深层结构如同种子萌芽(germinat-

ing seed)。然后,音乐仿佛有了自己的意志(volition),如同生命体一般,遵循展开/延长的规则,苗壮成长为参天大树一般的前景。高点的后移以至于时间箭头,就是源于申克分析中,展开/延长对于深层结构开头部分的偏爱(favor)。这一偏爱,直观体现在了申克图的一个特征:背景层中的大部分音符都聚集在曲目的后半段,甚至是结尾(见谱例 12)。

谱例 12　Salzer 在《结构化聆听》中所作的申克图谱,分析舒伯特的《圆舞曲》,Op. 18, No. 10

　　唯一指向性条件的案例:AB-AAB 左偏向模型

　　综合上述,我们试图构造一个形式主义的描述,用以精准捕获何谓"偏爱高点前的部分"。

　　AB-AAB 左偏向模型:我们从任意的二段音乐结构"AB"作为最小的曲式单位开始。时间箭头的存在,让我们在扩展 AB 这一结构的时候,偏爱前置加入内容(prepending)而不是后置(ap-

pending)。因此,我们倾向于在 AB 左侧加入内容。此外,我们倾向于复制 A 段,得到左侧复制(left-replication):AB↦AAB

　　AB-AAB 左偏向模型很好归纳了音乐中很多看似无关的现象,并为这些现象找到了一个共同的源头:时间箭头。我们在这里枚举几个例子:

（1）巴尔形式（bar form）

AAB 结构的曲式，往往又被称作"巴尔形式"。正因如此，巴尔形式可以被看作是"AB-AAB 左偏向模型"字面意义上最直观的例子。在《新格罗夫音乐与音乐家辞典》的词条中，巴尔形式被称作"最常见的音乐曲式之一"。巴尔形式所描述的 AAB 结构广泛存在于音乐中，也能凸显"AB-AAB 左偏向模型"的解释力。同时，巴尔形式作为本文讨论的案例，进一步证实了本文讨论的时间箭头：AAB 的曲式常见，但它的时间反演（time-reversal）对应形式，即 ABB，并不是一个

常见的曲式。这一时间反演对应形式的缺失，跟前文所提到的观点如出一辙："对过去的记录"没有面向未来的对应概念。

（2）句式（phrase model）

巴尔形式的 AAB，并不限于曲式，而是可以存在于任何的结构层级（hierarchical level）上。因此，我们可以试提出：句式中的"句子结构"（sentence structure）因为其 AAB 结构，可以被视作乐句层面的巴尔形式。如图 14 所示，句子结构的特征除了"短-短-长"（往往是 1 : 1 : 2 的比例）的节奏分组[35]，还有 AAB 的乐句内容划分。

谱例13　舒曼《幽灵变奏曲》，WoO24，开头主题部分[36]

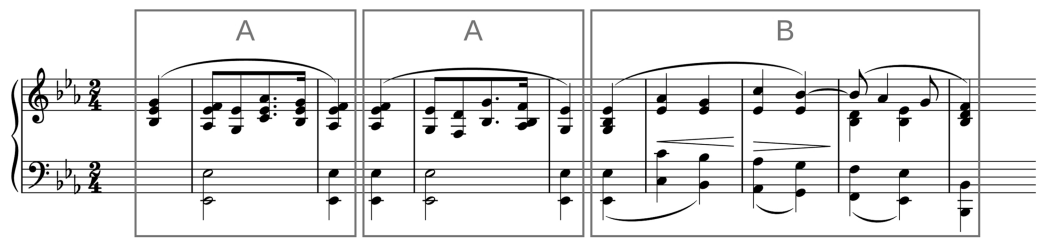

（3）申克分析中的"阻断"（interruption）

在申克分析中，一个深层结构可以在进行中途，到达属和声之后被阻断（这一概念得名于此），并且回到这一深层结构的开端重新开始（见谱例 14）。与其将其看作一个对和声进行的阻断，"AB-AAB 左偏向模型"提供了另一个视角（见谱例 15）：将深层结构标注为 AB，那么"阻断"这一概念无非就是"AB-AAB 左偏向模型"的一个具体应用。

谱例14　申克分析中的"阻断"，由
图中的双实线表示[37]

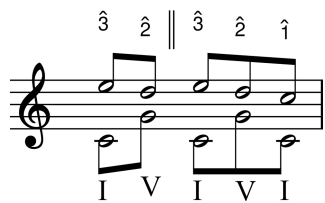

（4）奏鸣曲式中对呈示部的重复

根据习俗，我们往往用 ABA 来标记奏鸣曲式的曲式结构。然而，这一标记往往忽视了奏鸣曲中呈示部的重复。这一忽视凸显了音乐分析过度依赖符号系统的潜在弊端：重复记号因为不占用过多

谱例15　申克分析中的"阻断"[38]

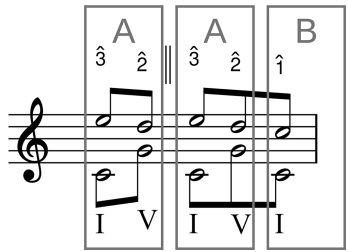

谱面空间，往往在分析中被忽视。相比之下，演奏者因为需要耗费时间来奏出呈示部的重复，会不可避免地重视呈示部的重复（甚至相比其他段落更加重视，因为需要考虑如何在段落的重复中保持音乐表现力）。在著作《奏鸣曲理论的元素》（Elements of Sonata Theory）中，作者明确指出了段落（例如呈示部）重复的重要性，不应在分析中被忽视。沿袭这一观念，我们在标记奏鸣曲式时，可以将呈示部的重复看作曲式结构的一部分，并因此得到 AABA 的标记。也就是说，奏鸣曲式中加入对呈示部的重复，并且把重复纳入曲式分析中，就是"AB-AAB 左偏向模型"的一个具体应用。此外，"AB-AAB 左偏

向模型"配合对呈示部重复的纳入分析,我们还可以重新解读奏鸣曲式和申克分析中"阻断"的关系。已有的一种观念认为,阻断对应的正是奏鸣曲中的再现部(见图2)。然而,结合上一段的讨论,我们可以进一步提出另一(alternative)解读:阻断对应的是呈示部的重复(见图3)。这一新解读为理解奏鸣曲式提供了一个新的视角,并可以潜在指导我们的演奏:例如,阻断往往给人一种音乐上"重置"(reset)的感觉,因为根据定义(by definition),音乐的深层结构被重置到了开头。那么,在知道了阻

| 申克分析 | $\hat{3}$ | $\hat{2}$ | $\hat{3}$ | $\hat{2}$ | $\hat{1}$ |
| | I | V | I | V | I |
| 奏鸣曲式 | 呈示部+发展部 | | 再现部 | | |

图2 示意图:申克分析中的"阻断"跟奏鸣曲式的对应关系

| 申克分析 | $\hat{3}$ | | $\hat{3}$ | $\hat{2}$ | | $\hat{1}$ |
| | I | | I | V | | I |
| 奏鸣曲式 | 呈示部 | 呈示部(重复) | | 发展部 | 再现部 |
| 曲式标记 | A | A | | B | A |

图3 示意图:申克分析中的"阻断"跟奏鸣曲式的对应关系[39]

断对应呈示部的重复,我们就可以在这一段落凸显出音乐"回到开始"的那种突兀截断。

通过上述例子的枚举我们可以看出,"AB-AAB左偏向模型"除了具有能够描述多个音乐现象所带来的解释力(explanatory power),还具有切实的指导意义。也就是说,这一模型不仅仅具有描述性(descriptive),还具有规定性(prescriptive),可以启发我们对音乐的理解和诠释。不论是这一模型,还是它所伴生的"高点后移",描述的都是一个动态的过程:从初始条件(initial condition)(AB曲式结构,未后移的高点)逐步演变到现状(AAB曲式结构,后移的高点)。它从诗意和现实的角度启发着我们:一方面,意识到了这一动态过程的初始条件的存在,我们所观测到的现状就不再是理所应当,而是来之不易;另一方面,这一来之不易,让我们可以把"AB-AAB左偏向模型","高点后移",以至于时间箭头所隐含的动态过程解读为诗意的斗争,那么AAB形式中生长出的A,则是雨后春笋,破茧成蝶,是羽化,更是新生。

(责任编辑:田可文)

# The Arrow of Time in Music
## ──Revisiting the Temporal Structure of Music with Distinguishability and Unique Orientability as the Anchor Point

Xu Qi

**Abstract**:Driven by the term "the arrow of time" as a general topic, the article develops a musical discussion by referring to the etymological origin of the term: philosophy (epistemology) and physics (thermodynamics). In particular, the article explores two specific conditions: distinguishability and unique orientability, from which the article derives respective musical propositions and case studies. For the distinguishability condition, the article focuses on the "recurrence" in music and tries to interpret Bach's *Christmas Oratorio* from the perspective of "birth/resurrection". For the unique orientability condition, the article discusses the process of delaying the climax, thereby proposing "AB-AAB left-replication" model, implying an organicist view by treating the temporal structure of music (e. g. form) as the product of a dynamic process: organic growth.

**Key words**:the arrow of time; organicism; Schenkerian analysis; epistemology; thermodynamics; sonata form

---

㉟图3是图2的另一解读,考虑到了其呈示部的重复问题。